\documentclass[12pt,preprint]{aastex}    

\newcommand \eg     {{\it e.g., }}

\newcommand \ie     {{\it i.e.,}}

\newcommand \eq     {\,=\,}                 

\newcommand \sinc   {{\rm{sinc}}}
\newcommand\convolution {{ * }}

\begin{document}

\title{Tip-tilt Error in Lyot Coronagraphs}

\author{James P. Lloyd \altaffilmark{1,2,3}}
	\affil{Astronomy Department\\
		California Institute of Technology\\
		1200 East California Boulevard, 
		Pasadena, CA 91125}
		\and
\author{Anand Sivaramakrishnan \altaffilmark{2}}
	\affil{Space Telescope Science Institute\\
		3700 San Martin Drive, Baltimore, MD 21218}
	%
\altaffiltext{1}{Millikan Fellow}
\altaffiltext{2}{NSF Center for Adaptive Optics}
\altaffiltext{3}{Present Address: Department of Astronomy, Cornell University, Ithaca NY}
\begin{abstract} 

The direct detection of extrasolar planets by imaging means is 
limited by the large flux of light from the host star being scattered
into the region of interest by a variety of processes, including
diffraction.
Coronagraphs are devices that suppress the undesirable scattering of
light caused by diffraction.  In a coronagraph  the sensitivity limit
for high dynamic range is limited by the propagation of errors
introduced by the imperfect optical system to the final image.  In
this paper we develop theory and simulations to understand how such
errors propagate in a coronagraph.  We describe the response of
classical and band-limited Lyot coronagraphs to small and large errors
in the placement of the central star, and identify ways of making such
coronagraphs more robust to small guiding errors.  We also uncover
features of the decentered PSF that can lead to spurious detection of
companions, especially with aggressive, high dynamic range
coronagraphs dedicated to companion searches aimed at finding
extrasolar terrestrial or Jovian planets.
\end{abstract}



\section{Introduction}

It is undesirable that the effect of edge diffraction from the entrance
aperture of a telescope results in the scattering of light into regions
of great interest for study of the circumstellar environment of stars.
The purpose of 
 coronagraphs is to select or modify the spatial frequency
content of the light, to effect suppression of diffracted light in 
a desired manner.  Discussion of the theory of diffraction limited 
stellar coronagraphs is typically limited to the on-axis point spread function
(PSF), often with the assumption of perfect optics.  In recent years
there has been an explosion of new concepts for coronagraphs, many of
which can achieve contrasts of $10^{-10}$ appropriate for Terrestrial
Planet Finder (TPF) applications in the absence of phase errors.  
A key question is the tolerance of coronagraphs to the variety
of imperfections that might be encountered in the real world.
Some previous studies have incorporated models of
phase errors (e.g. \citet{Malbet96,Sivaramakrishnan01,Green03spie}), although
these have not focused 
on delivering insight into how the errors propagate to the final image,
and how to design a more robust coronagraph at a conceptual level.

Here we focus our initial analysis on the propagation of tip-tilt
errors in Lyot coronagraphs \citep{Lyot39}.  A coronagraph is 
an instrument that suppresses light in a specific position in image
space, and thus has a spatially variable PSF.  The connection between 
tip-tilt of the wavefront (or equivalently decenter of the focal plane stop) 
and the response of the final image plane is important as an error
source, and leads to fundamental insight into, and understanding of, the 
operation of a Lyot coronagraph.

A hard-edged (binary) Lyot coronagraph is remarkably tolerant of 
tip-tilt errors, even for very small focal spots.  This is curious,
particularly given that one of the most scientifically successful 
coronagraphs, the Johns Hopkins Adaptive Optics Coronagraph (AOC) 
\citep{Golimowski92} responsible for the discovery of the 
first bona fide brown dwarf \citep{Nakajima94} incorporated a tip-tilt
system.  Whilst it was envisioned that this would improve the performance
of the coronagraph, in fact there is little benefit to suppression of
diffracted light, as shown below.  Also surprising is the 
counter-intuitive result that a graded or apodized focal plane
spot is {\em less} sensitive to small tip-tilt errors than a hard-edged
coronagraph, despite the fact that more light passes through the 
partially transmissive stop when the star wanders off axis.

\section{Second order monochromatic coronagraphic theory}

\begin{figure}
\epsscale{.5}
\plotone{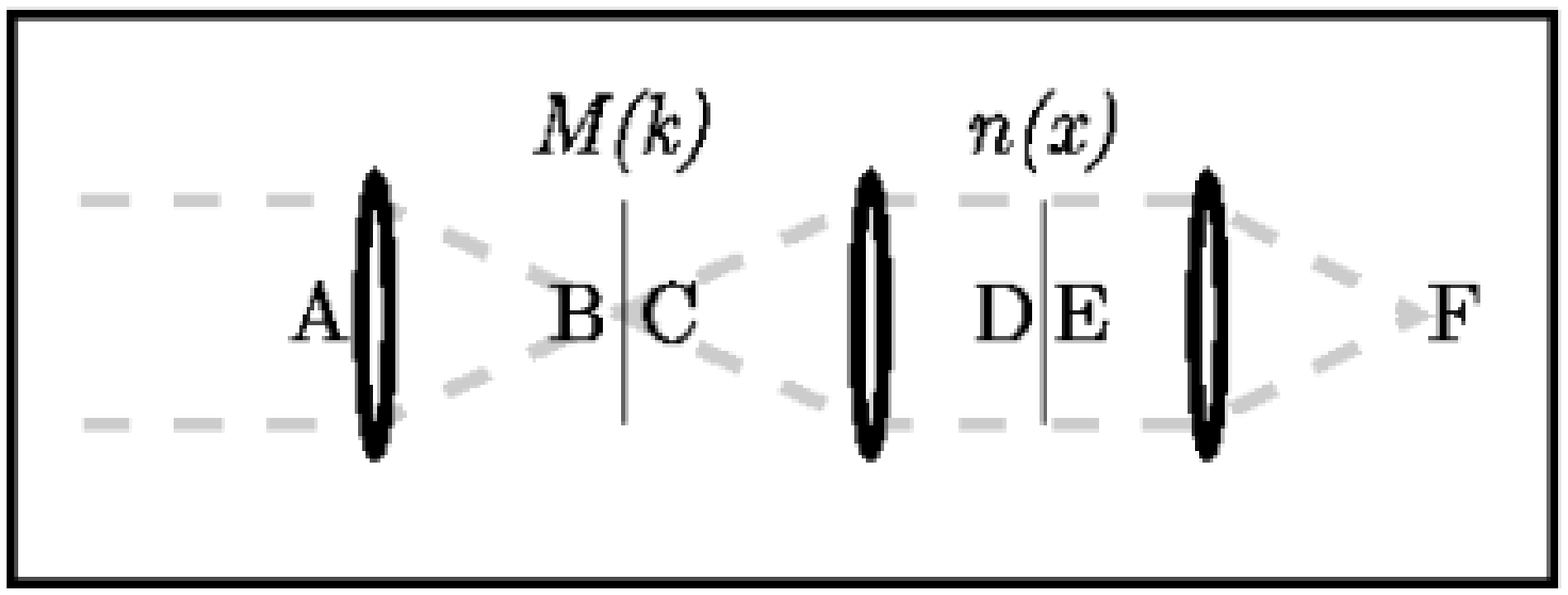}
\caption{
The essential planes and stops in a coronagraph.  The entrance aperture
is A, the direct image at B falls on a mask whose transmission function
is $M(k)$. The re-imaged pupil plane D, after being modified by passage through
a Lyot stop with a transmission function $n(x)$, is sent to the coronagraphic
image at F.
} 
\label{fig:corolayout}
\end{figure}

The phase on the telescope aperture is
$\phi(x)$, where $x=(x_1,x_2)$ is the location in
the aperture, in units of the wavelength of the light
(see Figure~\ref{fig:corolayout}).
The corresponding aperture illumination function describing
the electric field amplitude and relative phase in the pupil is
						$
		E_A =  A(x) \, e^{i\phi(x)}    
		    =  A(x) \, (1 + i\phi(x)  -  \phi(x)^2/2 + ...)
                        $, 
whose Fourier transform, 
						$
		E_B = a(k) \convolution (\delta(k) + i\Phi(k)  -   
		      \Phi(k) \convolution \Phi(k)/2 + ...)
                        $, 
is the electric field in the image plane B.
$\delta$ is the two-dimensional Dirac delta function,
and $k=(k_1,k_2)$ is
the image plane coordinate, in radians and $\convolution$ is the convolution operator.
Our convention is to change the case of a function to indicate its Fourier
transform.
\placefigure{fig:corolayout}
We multiply  the image field $E_B$ by a mask function $M(k)$ to
model the image plane stop of the coronagraph.
The image field immediately after the stop is $ E_C = M(k) \,  E_B $.
The electric field in the re-imaged pupil following the 
image plane stop, $E_D$, is the Fourier transform of $E_C$. 
We use the fact that the transform of $E_B$ is just the aperture
illumination function $E_A$ itself:
						\begin{eqnarray} \label{ED} 
		E_D &=&  m(x) \convolution  E_A    \nonumber \\
		    &=&  m(x) \convolution [ A(x) \, (1 + i\phi(x)  -  \nonumber  \\
			&&                            \phi(x)^2/2 + ... )]
                        \end{eqnarray}
If the Lyot pupil stop transmission is $n(x)$, the electric
field after the Lyot stop is $E_E = n(x) E_D$.  The transform of this
expression is the final coronagraphic image field strength
when the wavefront phasor is expanded as a power series in the phase $\phi$:
						\begin{eqnarray} \label{EF} 
		E_F &=& N(k) \convolution [M(k) \, E_B  ]  \nonumber \\
		    &=& N(k) \convolution [M(k) \, (a(k) \convolution (\delta(k) + i\Phi(k)  - \nonumber \\
			 &&                                  \Phi(k) \convolution \Phi(k)/2 + ...))].
                        \end{eqnarray}
Understanding high dynamic range Lyot coronagraphy hinges
on understanding the structure of the field strength $E_D$
in the Lyot plane located at D.

\section{Guiding error in a Lyot coronagraph} \label{LyotTT}

The effect of small tip-tilt errors on a Lyot coronagraph operating 
on a high Strehl ratio image is described by 
a truncated version of equation (\ref{ED}).
The mask function in a Lyot coronagraph is best expressed as
$M(k) = 1 - W(k)$, where $W(k)$ is the `image stop shape'
function.  For a hard-edged stop $W(k) = \Pi(Dk/s)$, where $s$
is the image stop diameter in units of the resolution of the 
optical system.  If the image plane stop is opaque at its center, 
$W(0) = 1$ (which constrains $w(x)$ to have unit area).
The FT of the stop function $M(k)$ is $m(x) = \delta(x) - w(x)$,
so the Lyot pupil electric field of a Lyot coronagraph
can be expressed as
						\begin{eqnarray} \label{EDL} 
		E_D &=&  [\delta(x) - w(x)]  \convolution [ A(x) \, (1 + i\phi(x)  -  \nonumber  \\
			&&                            \phi(x)^2/2)]
                        \end{eqnarray}
for sufficiently small phase errors (\ie\ $|\phi| \ll 1$) in the pupil.
Pure tip-tilt error is described by a phase function 
$\phi(x) = \alpha \cdot x \equiv \alpha_1 x_1 + \alpha_2 x_2$
($\alpha$ is in radians per wavelength in pupil space).
We require that the image displacement be much less than a diffraction
width, so $|\alpha| D \ll 1$.  
Following the method developed in \citet{Sivaramakrishnan02}, and truncating our 
expansion at the second order, we derive
an analytical expression for the Lyot pupil field (which
is typically valid for Strehl ratios of the order of 95\% and above
\citep{Perrin03}):
						\begin{eqnarray} \label{EDLlin} 
		E_D &=&  [\delta(x) - w(x)]  \convolution [ A(x) \, (1 + i\alpha \cdot x  -  \nonumber \\
		    &&                                    (\alpha \cdot x)^2/2 + ... )].  
                        \end{eqnarray}
$E_D$ is therefore the sum of a zero order term
						\begin{equation} \label{EL0}
		E_{L0} \eq A(x) - w(x) \convolution A(x),
						\end{equation}
a first order term
						\begin{equation} \label{EL1}
		E_{L1} \eq i [\alpha \cdot x A(x) - w(x) \convolution (\alpha \cdot x A(x)], 
						\end{equation}
and a second order term
						\begin{equation} \label{EL2}
		E_{L2} \eq - \onehalf [(\alpha \cdot x)^2 A(x) - w(x) \convolution ((\alpha \cdot x)^2 A(x))].
						\end{equation}

The behavior of these three terms is most easily understood by
following this analysis in the case of a band-limited Lyot coronagraph
\citep{Kuchner02}.  We use a coronagraph with an image plane stop
shape function which possesses a FT of $w(x) = \Pi(x_1/\epsilon,
x_2/\epsilon) / \epsilon^2$, where $\epsilon = D/s$ ($s$ is of the
order of a few to $10$, and corresponds to the `size' of the image
plane stop in units of $\lambda/D$).  This simplifies the analytical
calculations and brings out the salient features of the manner in
which tilt errors propagate through a Lyot coronagraph.  For a 
hard-edged focal stop, $w(x)$ is a sinc function (see Figures~\ref{fig:1dcoro} and \ref{fig:1dcoro-tilt}).  Once we are
armed with a theoretical understanding of the expressions in equations
(\ref{EL0}), (\ref{EL1}), and (\ref{EL2}), we can investigate the
response of more common Lyot coronagraph designs to guiding errors
numerically, and also start to address how pupil apodization affects
the way guiding errors degrade dynamic range.

The zero order term is well-understood for Lyot coronagraphs 
(\eg\ \citet{Sivaramakrishnan01} and references therein), and is outlined in 
Figure~\ref{fig:1dcoro}.
                    %
\placefigure{fig:1dcoro}
\placefigure{fig:1dcoro-tilt}

\subsection{First order tip-tilt leak } \label{LyotTT1}

The first order term allows light through only at the edges of
an unapodized pupil.  Such behavior is 
similar to the zero order term.  The leaked light can be suppressed 
by the usual undersizing of the Lyot stop.
In order to see why this is true,
one must consider the value of the convolution of the `small' 
two-dimensional unit-area top-hat function $w(x)$ with the
function $xA(x)$,
as shown in Figure~\ref{fig:1dcoro-tilt}.
Let us consider an x-tilt (by setting $\alpha_2 = 0$).
If $A(x) = 1$ inside the pupil, $xA(x)$ is a flat plane with
slope $\alpha_1$ passing through the origin, and containing the $x_2$ axis.
The value of the convolution integral when the top-hat function lies
entirely within the support of the aperture is simply the $x_1$ value of
the offset.  Therefore in the interior of the pupil
$xA(x) = w(x) \convolution xA(x) = x_1$. The electric field further than $\epsilon = D/s$ from the pupil boundary is zero.

A hard-edged focal stop results in leakage of light into the interior of 
the pupil from the wings of the sinc function (see Figure~\ref{fig:1dcoro-tilt} ).
A graded focal stop has a more compact Fourier transform than a hard-edged stop.
In this case the interior of the Lyot pupil, where the field remains zero,
is larger than that of a hard-edged coronagraph's Lyot pupil.  This
results in less sensitivity to tilt error for the same Lyot plane
stop geometry at high Strehl ratios, even when the tilt errors are 
large enough to move the star into regions of the focal stop with significant
transmission.

\subsection{Second order tip-tilt leak through} \label{LyotTT2}
We apply similar logic to the second order term.  In the special case
of a clear pupil, and the same band-limited coronagraph,
the Lyot pupil electric field depends on the difference between 
  $x_1^2 A(x)$ and $w(x) \convolution (x_1^2 A(x))$.
The convolution integral is no longer the identity operator
even when the top-hat function lies entirely within the pupil
support.  There is a uniform residual field strength approximately
equal to $\alpha_1^2 \epsilon^2/8$ everywhere in the interior.
There is also the same ``bright edge'' effect as is seen in the
zero and first order terms, but that is removed by the optimally undersized
Lyot stop.  The uniform background in the pupil plane from the
second order contribution of a pure tilt term causes a 
``ghostly PSF'' to form on axis (not displaced) even with an optimized
Lyot stop (see e.g. Figure~\ref{fig:flc}).  The energy in this PSF varies as
the fourth power of the (small) tilt error, and inversely as the
fourth power of the focal plane stop diameter.  First order effects of defocus will affect 
the coronagraph in a similar way.  It is the combination of these
``ghostly PSFs'' with the real PSF of the star that results in the 
distorted images shown in Figure~\ref{fig:flc}.

\section{The Point-Spread Functions of a Lyot Coronagraph} \label{LyotTTlarge}
\begin{figure}
\epsscale{0.75}
\plotone{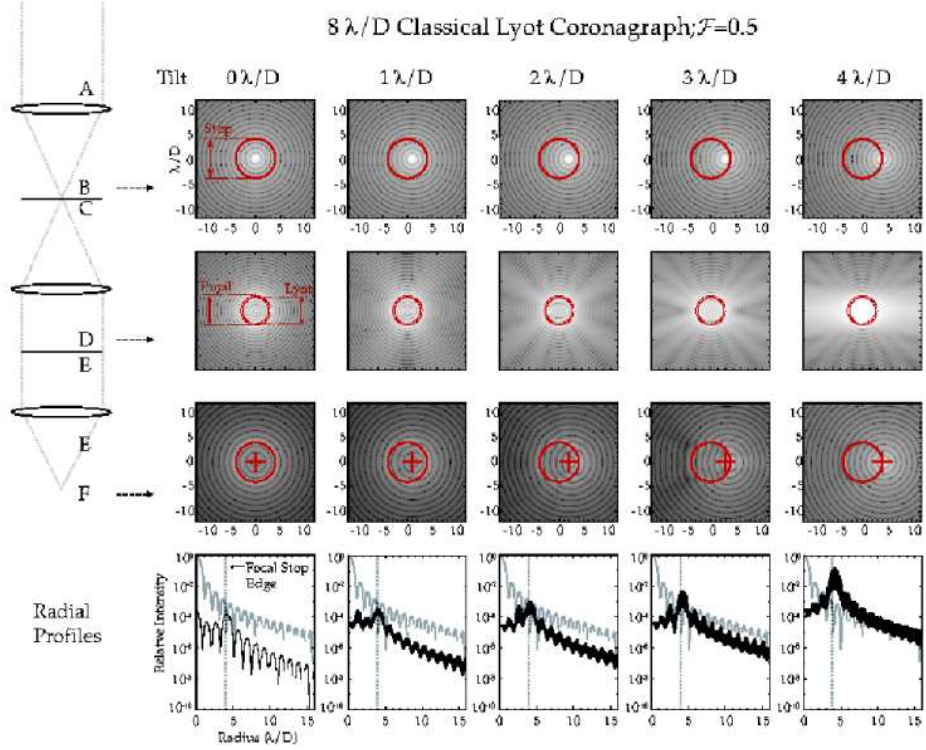}
\caption{
Focal, Lyot plane and final coronagraphic image intensities for a 
hard-edged Lyot 
coronagraph with varying degrees of tilt.  The occulting spot is
$8\lambda/D$ diameter, and outlined in red in the focal plane images.
The outline of the pupil and undersized Lyot stop are shown in red 
in the Lyot plane images.  The outline of the image of the focal stop
is shown in red in the final coronagraphic image.  In the final coronagraphic
image, the position of the star in the image plane is marked with a cross.  
If the star is behind the focal stop, the peak in the coronagraphic
image does not correspond to the position of the star, leading to 
`fake sources'.   The radial profiles show the range from the mean to 
maximum intensity in an annulus centered on the center of the focal stop.
The non-coronagraphic Airy pattern is shown for comparison.
}
\label{fig:flc}
\end{figure}

\placefigure{fig:flc}

Up to this point we have concerned ourselves with small ($\ll \lambda/D$)
tip-tilt errors in Lyot coronagraphs.  Here we lift that constraint, and
examine the morphology of the PSF of a Lyot coronagraph over a wide
range of stellar locations relative to the spot center.

We simulated the PSF of a coronagraph when a star is offset from the
center of the stop.  These PSFs are illustrated in
Figure~\ref{fig:flc} using a spot $8\lambda/D$ in diameter, although
we studied both smaller and larger stops.  We found markedly different
morphologies in three regimes.  When the star behind the spot is
displaced a small amount, the PSF looks similar to that of the
perfectly aligned coronagraph.  The rows in Figure~\ref{fig:flc}
show a sequence of locations of the central star, beginning at the
very center of the occulting spot, with a Lyot stop
diameter 75\% of the entrance aperture diameter.  When the star is within
$\lambda/D$ of the spot edge, the PSF develops outcrops that are not at the location of the star.  When the star is located at the very edge of
the spot, or outside it, the PSF takes on a typical direct image PSF
shape.

The three rows of images in Figure~\ref{fig:flc} are the
PSF in the first focal plane, the Lyot pupil plane intensity,
and the final coronagraphic PSF and shown in radial profile.
We note the appearance of the fake source located about $2\lambda/D$ 
from the star in the coronagraphic PSF at a misalignment of $2\lambda/D$.
The manner in which placement errors interact with higher order
errors, such as spherical aberration, has not been studied yet.
This suggests that PSF modeling of coronagraphic data should be
performed with care to avoid misinterpreting structure close to the spot
edge in the image (e.g. \citet{Krist98,Krist04}).

This exercise is relevant to coronagraphy on very high Strehl ratio images,
although it also has immediate applicability to coronagraphic science carried
out today, with the HST ACS, for instance, if bright structures were present
behind the focal stop but near its edge.

\begin{figure}
\plotone{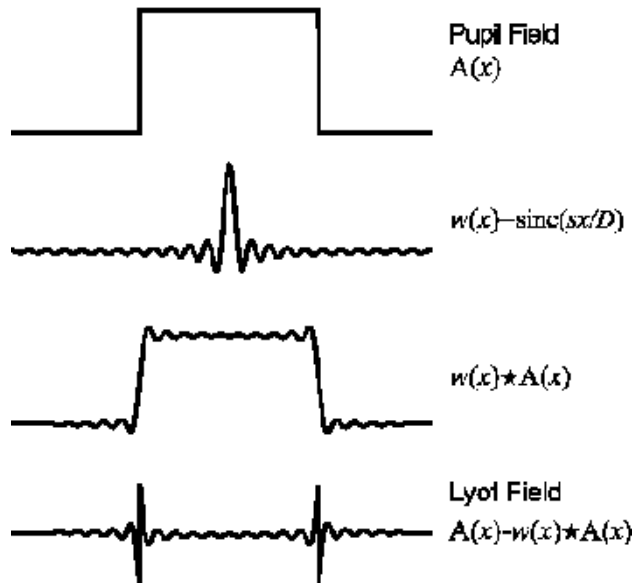}
\caption{ One dimensional representation of a perfectly aligned 
hard-edged Lyot coronagraph.  A band limited stop with a top-hat function
bandpass does not have the ringing in the wings of the sinc function.
There is no fundamental difference between these designs for the purposes here,
since $w(x)$ has approximately the same spatial scale for both.
Compared to a hard-edged stop, apodizing the focal stop
reduces the ringing in the sinc function, resulting in less light
bleeding into the center of the pupil.  }
\label{fig:1dcoro}
\end{figure}

\begin{figure}
\plotone{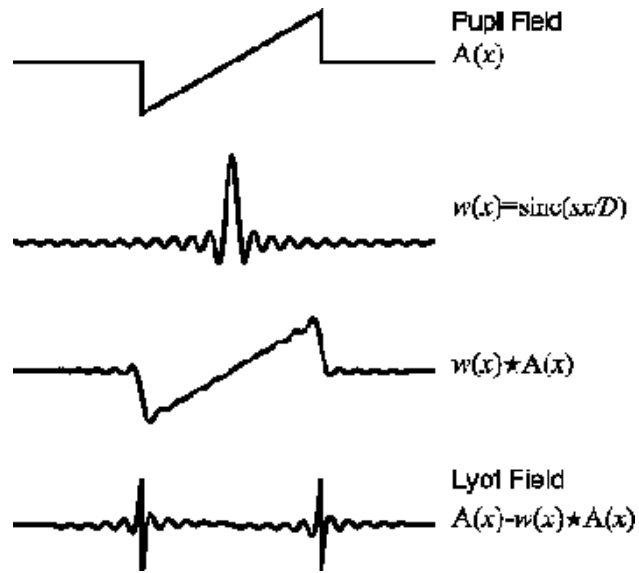}
\caption{
One dimensional representation of the first-order leak due to tilt error
in a Lyot coronagraph (see equation (\ref{EL1}).  The 
effect of tilt is largely confined to the edge of the pupil, which is
already suppressed by an optimized Lyot stop.
} 
\label{fig:1dcoro-tilt}
\end{figure}

\begin{figure}
\epsscale{.6}
\plotone{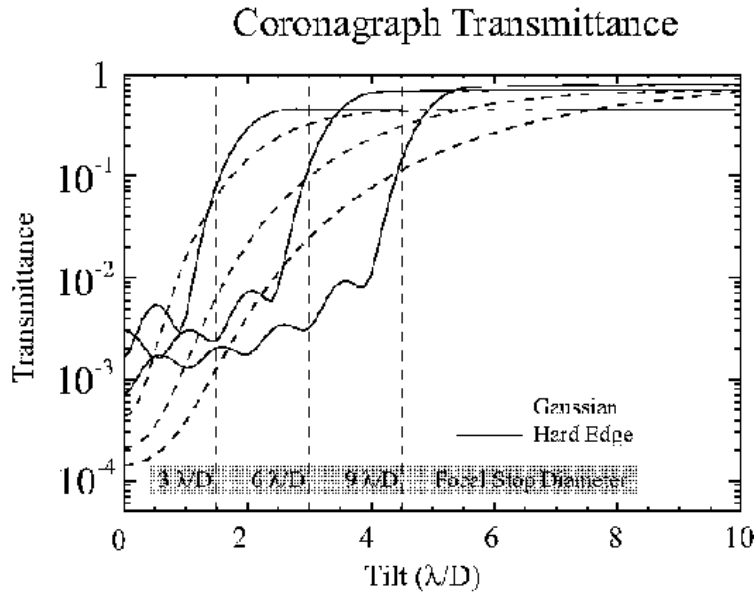}
\caption{ Coronagraph transmittance as a function of tilt for three 
hard-edged and Gaussian focal stop diameters.  Transmittance is defined as
the fraction of light entering the system aperture that propagates to
the final image plane outside the image of the focal stop.  It is the total
fraction of light that the coronagraph suppresses, not the
on-axis null depth.  Note that the suppression of the 6 $\lambda$/D
hard-edge coronagraph improves with small tilt errors as a
result of the phasing of the dark/bright Airy pattern with respect to
the stop edge.  The rejection factor asymptotes to the fractional
throughput of the ${\cal F}=0.5$ undersized Lyot stop (of diameter $(1-1/s)D$ for an
$s\lambda/D$ focal stop), which always blocks a fraction of the light.
}
\label{fig:leakplot}
\end{figure}

\begin{figure}
\plotone{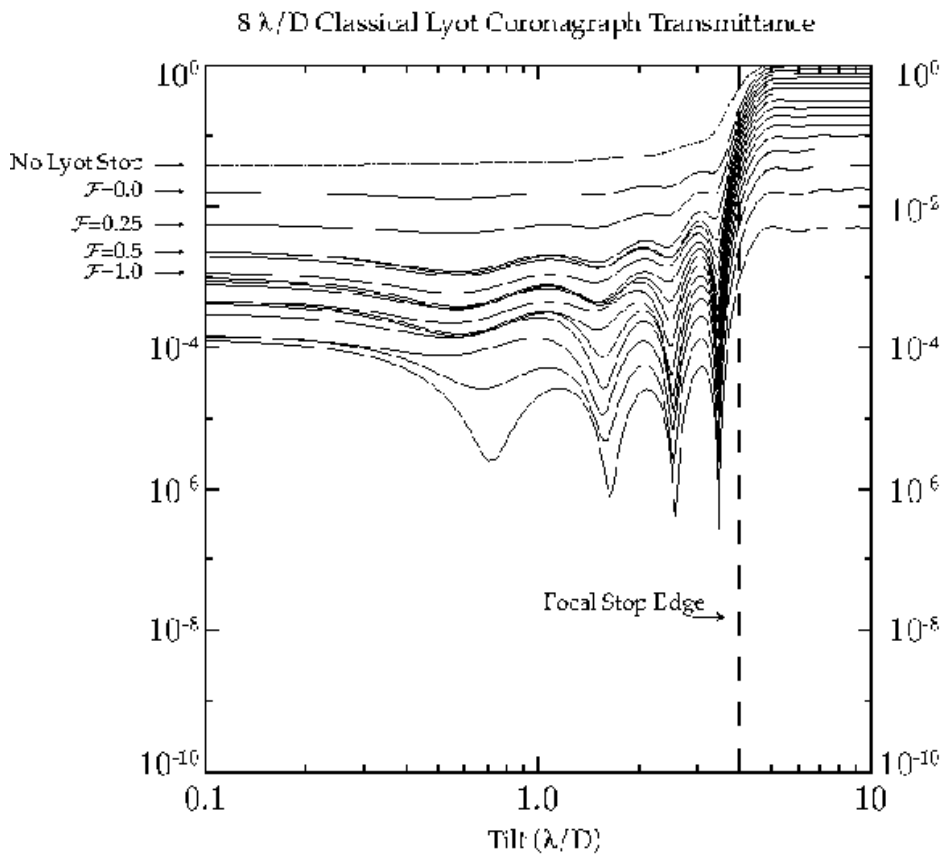}
\caption{
Combined effect of Lyot stop tuning parameter ${\cal F}$ and tilt errors on a 
hard-edged Lyot corongraph indicated by total coronagraph transmittance.  A family of Lyot coronagraphs with 
8 $\lambda/D$ diameter focal stop and varying Lyot stop diameters is shown.  
The no Lyot stop case accounts only for the fraction of energy suppressed
by the focal stop.  The progressive undersizing of the Lyot stop from ${\cal F}=0$ 
(a Lyot stop that is the exact image of the input pupil) in steps of ${\cal F}=0.25$ 
rejects both on-axis and off-axis light.  The point of diminishing returns is at ${\cal F}\sim 0.5$
as found by \citet{Sivaramakrishnan01}.  The transmittance asymptotes to the 
transmission of the Lyot stop.
}
\label{fig:hardtuningplot}
\end{figure}

Figure~\ref{fig:leakplot} shows coronagraphic rejection efficiency 
as a function of tilt error for several focal plane stops.
Typical coronagraphic reductions of the best current
space-based data demonstrate that imperfect calibration data and
temporal variations in the PSF set the limits on dynamic range
\citep{Krist98}, so we avoid using simplistic estimates of dynamic
range using monochromatic simulations to evaluate the actual effects
of tilt errors.
We use the fraction of transmitted central source light as a metric of
coronagraphic performance.  We define the transmittance of a
coronagraph to be the integrated light in the final focal plane, excluding the
region inside the focal stop (weighted by the focal stop transmission for the Gaussian case).
This quantity is directly related to the
photon-limited noise, albeit qualitatively.  
We choose this quantity as a metric for the purposes of this paper in preference to contrast, as it is independent of
the choice of an inner working angle for the coronagraph.  The transmittance of
light is calculated for $s=$ 3, 6, and 9 $\lambda/D$ focal stops, with
a hard-edged Lyot stop undersized by $0.5 \times 1/s$ of the pupil radius.  A
Gaussian apodized focal stop is compared to the hard-edged focal stop.
Since the Fourier transform of an apodized stop is more compact, less
light bleeds into the center of the pupil (see Figures~\ref{fig:1dcoro} and \ref{fig:1dcoro-tilt}).  It
is remarkable that the Gaussian apodized stop is more efficient even
in the presence of quite large tilt errors.  For example, a
$3\lambda/D$ FWHM Gaussian focal stop suppresses more light than a
$3\lambda/D$ diameter hard-edged stop even at $0.7 \lambda/D$ tilt
(see Figure~\ref{fig:leakplot}), despite the transmission of the stop being $10\%$ at
this radius.  %
The remarkable robustness of the classical Lyot coronagraph is
apparent in Figures~\ref{fig:flc} and \ref{fig:hardtuningplot}, contrary to the expectation 
that led \citet{Golimowski92} and \citet{Lloyd01} to incorporated tip/tilt control systems
into Lyot coronagraphs.  The leakage of light from the central star
 remains concentrated close to the edge of
the image of the focal stop until the central star gets
to within a resolution element of the stop edge.  This fact, combined
with its ease of manufacture and its broad-band performance, makes the
Lyot coronagraph interesting even in the era of novel coronagraphic
designs, which must all be well-understood in terms of tolerance to the
variety of errors that might exist in real telescopes.

\begin{figure}
\plotone{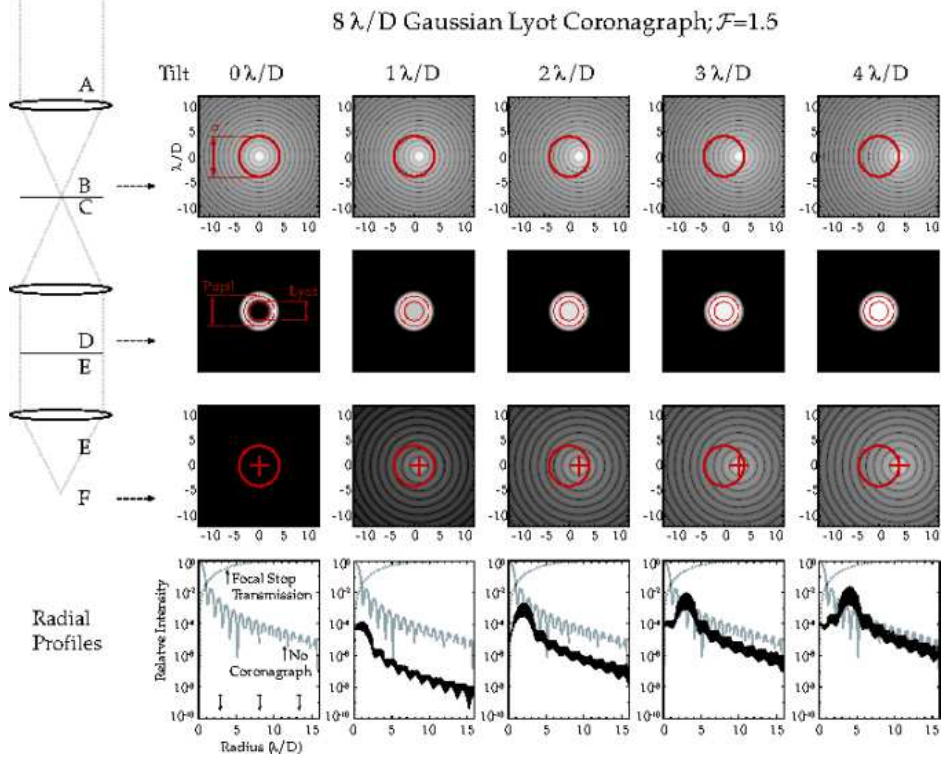}
\caption{
Focal, Lyot plane and final coronagraphic image intensities for a 
Gaussian Lyot 
coronagraph with varying degrees of tilt.  The occulting mask  is
$\sigma = 8\lambda/D$ diameter, and outlined in red in the focal plane images.
The outline of the pupil and undersized Lyot stop are shown in red 
in the Lyot plane images.  The Lyot stop is undersized more aggressively
than in the case of a hard-edged coronagraph (see Figure 8 and discussion in 
the text).  The outline of the image of the focal stop
is shown in red in the final coronagraphic image.  In the final coronagraphic
image, the position of the star in the image plane is marked with a cross.  
The radial profiles show the range from the mean to 
maximum intensity in an annulus centered on the center of the focal stop.
The non-coronagraphic Airy pattern and the transmission profile 
of the mask are shown for comparison.  Note that the suppression of 
diffraction in the wings is superior to the hard-edged case, even for 
tilt errors up to 3$\lambda/D$, at which the Gaussian stop transmission is 7\%.
}
\label{fig:flcgauss}
\end{figure}

\begin{figure}
\plotone{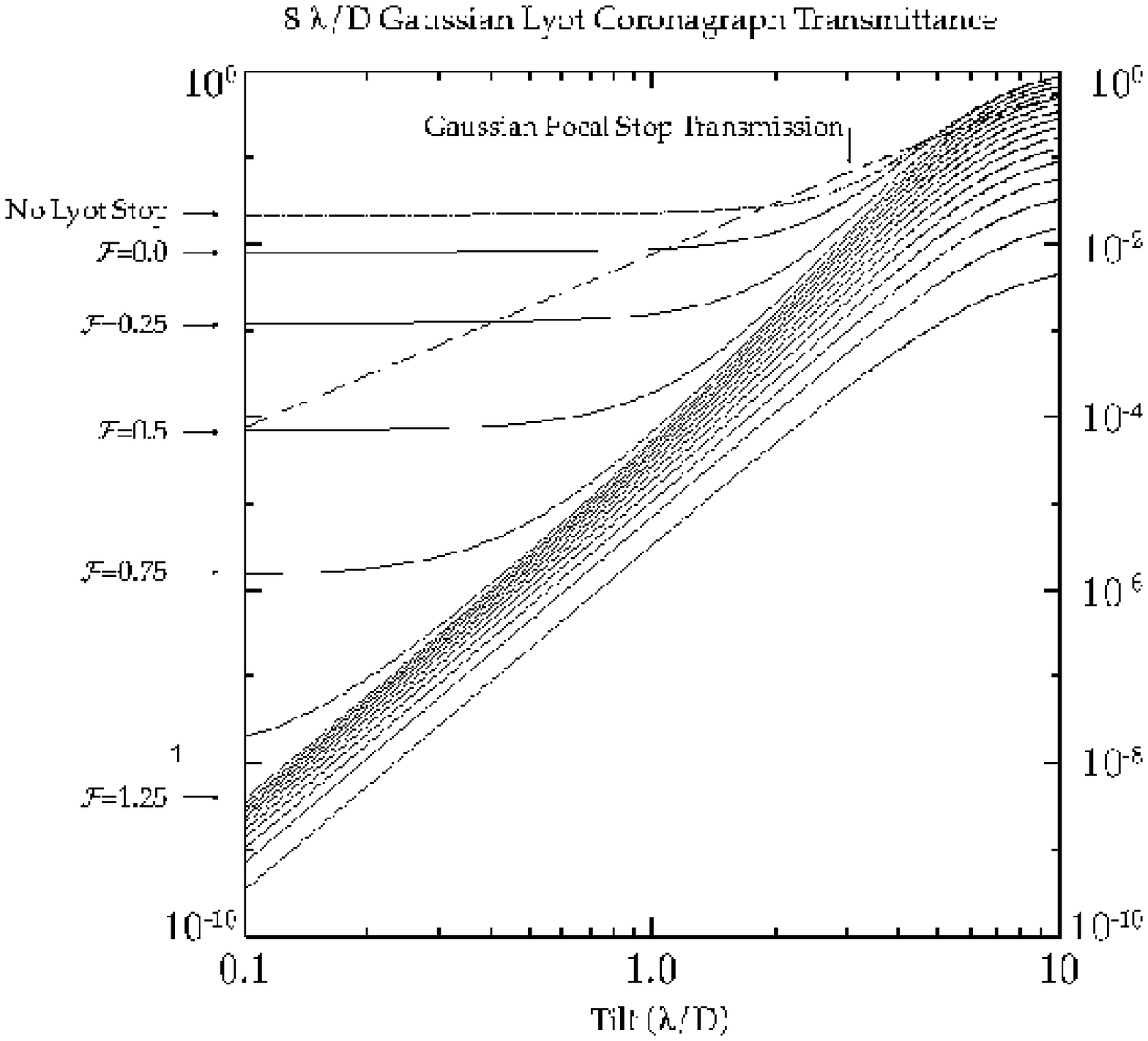}
\caption{
Combined effect of Lyot stop tuning parameter ${\cal F}$ and tilt errors on a 
Gaussian Lyot corongraph indicated by total coronagraph transmittance.  A family of
Gaussian  Lyot coronagraphs with 
$\sigma=8\lambda/D$ focal stop and varying Lyot stop diameters is shown.  
The no Lyot stop case accounts only for the fraction of energy suppressed
by the focal stop.  The progressive undersizing of the Lyot stop from ${\cal F}=0$ 
(a Lyot stop that is the exact image of the input pupil) in steps of ${\cal F}=0.25$ 
rejects both on-axis and off-axis light.  Unlike the hard-edged case, the on-axis 
rejection continues to improve to ${\cal F}>1$ since Gaussian wings of a Gaussian continue to drop 
rapidly unlike the  wings of the \sinc function. The transmittance asymptotes to the 
transmission of the Lyot stop.
}
\label{fig:gausstuningplot}
\end{figure}

The comparison of Gaussian and hard-edge coronagraphs on an equal footing
is complicated by the definition of an appropriate equivalent width for the 
Gaussian stop, and the undersizing of the Lyot stop.  For the purposes of
comparison, we characterized the width of the Gaussian stop by $\sigma$ 
where the transmission of the stop is $1-\exp(-x^2/2\sigma^2)$.   
We adopt the convention  of \citet{Sivaramakrishnan01} and define a Lyot 
tuning paramater ${\cal F}$ which defines the fractional radial undersizing of the
Lyot stop in units of $D/s$ (or $D/\sigma$).   For a hard edges Lyot coronagraph,
${\cal F}\approx 0.5$ results in most of the performance benefits of undersizing the Lyot
stop, as the Lyot stop excludes the core of the $w(x)$ $\sinc$ function around the 
edge of the pupil.  Further undersizing in this case results in relatively small
gains as the wings of a sinc function decay slowly (this is calculated in detail in \citet{Makidon00}).  For a Gaussian stop, 
however, the wings are suppressed, and gains continue with further undersizing
(see Figure~\ref{fig:gausstuningplot}).  The ultimate application of such tapering of the focal stop 
to achieve the most compact $w(x)$ is the generalization to more arbitrary 
functions with the concept of the Band Limited Coronagraph \citep{Kuchner02}. 
The rejection of such coronagraphs continues to improve with extremely aggressive
undersizing of the Lyot stop (see Figure~\ref{fig:gausstuningplot}.   To achieve the very 
high contrast required for terrestrial planet detection a band-limited or gaussian coronagraph
with an aggressive Lyot stop (${\cal F} > 1.5$)  For such a coronagraph (see Figure~\ref{fig:flcgauss}), the near complete rejection of on-axis light is lost with even a small tilt error, but
the coronagraph remains robust against tip/tilt errors in the sense that the wings of 
the PSF are suppressed even for tip/tilt errors of a few $\lambda/D$.

\newpage

\acknowledgements

The authors wish to thank the Space Telescope Science Institute's
Research Programs Office, Visitor Program, and
Director's Discretionary Research Fund for support.
This work has also been supported by the National Science Foundation
Science and Technology Center for Adaptive Optics, managed by the
University of California at Santa Cruz under cooperative
agreement No.~AST-9876783, and 
 AFOSR and NSF jointly sponsored research under grants
AST-0335695 and AST-0334916.   We are indebted to Marc Kuchner 
and the anonymous referee for comments on this manuscript.

\bibliographystyle{apj}
\bibliography{ms}

\newpage


\end{document}